\documentclass[11pt]{article}
\usepackage{graphicx}
\usepackage{amsmath}
\usepackage{amssymb}
\usepackage{caption2}
\begin{document}
\bibliographystyle{prsty}
\begin{center}
{\large {\bf \sc{  Analysis of  the decay $B^0 \to \chi_{c1} \pi^0$  with light-cone QCD sum rules }}} \\[2mm]
Zhi-Gang Wang \footnote{E-mail,wangzgyiti@yahoo.com.cn.  }     \\
 Department of Physics, North China Electric Power University,
Baoding 071003, P. R. China

\end{center}

\begin{abstract}
In this article, we calculate  the contribution from the
nonfactorizable soft hadronic matrix element  to the  decay
$B^0\rightarrow \chi_{c1} \pi^0$ with  the light-cone QCD sum rules.
The numerical results show that its contribution is rather large and
should not be neglected. The total amplitudes lead to a branching
fraction which is in agreement with the experimental data
marginally.
\end{abstract}



\section{Introduction}

Recently, the Belle Collaboration   measured  the branching fraction
for the Cabibbo- and color-suppressed decay $B^0 \to \chi_{c1}\pi^0$
based on a data sample of $657\times 10^6$~$B\overline B$ events
collected at the $\Upsilon(4S)$ resonance with the Belle detector at
the KEKB asymmetric-energy $e^+e^-$ collider \cite{Belle0809}. The
signal is $40\pm9$ events with a significance of $4.7\sigma$
including systematic uncertainties, and the branching fraction is
about $ (1.12\pm 0.25\pm 0.12)\times 10^{-5}$.

The decay takes place through the process $b \to d c \bar{c}$ (or
more precise $\bar{b} \to \bar{d} c\bar{c}$, they relate with each
other by  charge conjunction, in this article, we calculate the
amplitudes for the process $b \to dc \bar{c}$, then take  charge
conjunction to obtain the branching fraction.) at the quark-level
\cite{Buras}.
 If the leading order tree diagram dominates, the time-dependent CP-violating
asymmetries  are predicted to be the same as the ones  in the $b \to
sc\bar c$ decays \cite{Sanda81}. The time-dependent CP-violation
parameters for the similar decay  $B^0 \to J/\psi \pi^0$ have been
measured by the Belle \cite{belle_psipi0-1,belle_psipi0-2} and Babar
\cite{babar_psipi0-1,babar_psipi0-2,babar_psipi0-3} Collaborations.
The deviation of the CP-violating asymmetries from those
expectations may indicate non-negligible contributions from the
penguin amplitudes or
 new physics.

 The quantitative understanding of the $B$-decays depends  on our knowledge about the
nonperturbative hadronic matrix elements of the operators entering
the effective weak Hamiltonian \cite{Buras}. In recent years, great
progresses have been made in this aspect, such as the generalized
factorization approach \cite{Ali1997,Cheng1999}, the QCD
factorization approach \cite{BBNS2000,BBNS2001}, the perturbative
QCD \cite{Li2001-1,Li2001-2}, the soft-collinear effective theory
\cite{Bauer2000}, etc.  Factorization of the hadronic matrix
elements has been proved to hold in the leading order  in many
processes.

The effects of the soft gluons which violate  factorization are
supposed of order $O(\frac{\Lambda_{QCD}}{m_b})$ and neglected in
the QCD-improved factorization studies \cite{BBNS2000,BBNS2001},
however, no theoretical work has ever proved that they are small
quantities. For the color-suppressed $B$ to charmonia decays, there
may be significant impacts of the nonfactorizable soft
contributions.

In Ref.\cite{Kh2001},  Khodjamirian introduce a  technique based on
the light-cone QCD sum rules to estimate the nonfactorizable soft
contributions,   where the soft gluon effects are represented by the
quark-antiquark-gluon distribution amplitudes of the light mesons,
the hadronic matrix element appears  as a part of the hadronic
dispersion relation for the correlation function. Thereafter, the
light-cone QCD sum rules are applied to study the nonfactorizable
hadronic matrix elements in the $B$-decays due to the soft gluons
exchanges \cite{Soft1,Soft2,Soft3,Soft4,Soft5,
Soft6,Soft7,Soft8,Soft9, Soft10,Soft11}.

 It is interesting to study the nonfactorizable soft
contributions in the  decay $B^0 \to \chi_{c1}\pi^0$ with the
light-cone QCD sum rules.

The article is organized as follows:   the factorizable
contributions from the effective weak Hamiltonian are derived in
Sec.2; the soft hadronic matrix element $\langle \chi_{c1} \pi^0|
{\cal \widetilde{O}}(0) | B \rangle $ is calculated with the
light-cone sum rules  approach in Sec.3;
 numerical results  are presented in Sec.4;
the section 5 is reserved for conclusion.

\section{Effective weak Hamiltonian and factorizable contributions}

The effective weak Hamiltonian for the $b\rightarrow d c \bar{c}$
decay modes can be written as (for detailed discussion of the
effective weak Hamiltonian, one can consult Ref. \cite{Buras})
\begin{equation}
H_w = \frac{G_F}{\sqrt{2}}\left\{V_{c b} V_{c d}^* \left[ C_1(\mu)
{\cal O}_1 +  C_2(\mu) {\cal O}_2 \right] -V_{tb}
V_{td}^*\sum_{i=3}^{10} C_i {\cal O}_i \right\}  \, ,
\end{equation}
where $V_{ij}$'s are the CKM matrix elements, $C_i$'s are the Wilson
coefficients calculated at the renormalization scale $\mu \sim
O(m_b)$ and the relevant operators ${\cal O}_i$  are given by
 \begin{eqnarray}
 {\cal O}_1&=&(\overline{d}_{\alpha} b_{\beta})_{V-A}
(\overline{c}_{\beta} c_{\alpha})_{V-A}\, , \nonumber\\
{\cal O}_2&=&(\overline{d}_{\alpha} b_{\alpha})_{V-A}
(\overline{c}_{\beta}
c_{\beta})_{V-A} \, ,  \nonumber\\
 {\cal O}_{3(5)}&=&(\overline{d}_{\alpha} b_{\alpha})_{V-A} \sum_q
(\overline{q}_{\beta} q_{\beta})_{V-A(V+A)}\, ,  \nonumber\\
 {\cal O}_{4(6)}&=&(\overline{d}_{\alpha} b_{\beta})_{V-A}  \sum_q
(\overline{q}_{\beta} q_{\alpha})_{V-A(V+A)} \, ,
 \end{eqnarray}
 where we have neglected the Wilson coefficients $C_7,C_8,C_9,C_{10}$
 due to their small values.  We can reorganize
the color-mismatched quark fields into color singlet states by Fierz
transformation,  and express the effective weak Hamiltonian $H_w$ in
the following form,
\begin{eqnarray}
H_w &=& \frac{G_F}{\sqrt{2}}   \left\{  V_{c b} V_{c d}^* \left[
\left ( C_2(\mu) + \frac{C_1(\mu)}{3} \right ) {\cal O}_2 +  2
C_1(\mu) {\cal \widetilde{O}}_2 \right]
 \right. \nonumber \\
&& -V_{tb} V_{td}^* \left[ \left ( C_3(\mu) + \frac{C_4(\mu)}{3}
\right ) {\cal O}_3 +  2 C_4(\mu) {\cal \widetilde{O}}_3 \right]
 \nonumber \\
&& \left. -V_{tb} V_{td}^* \left[ \left ( C_5(\mu) +
\frac{C_6(\mu)}{3} \right ) {\cal O}_5 +  2 C_6(\mu) {\cal
\widetilde{O}}_5 \right] \right\} \, ,
\end{eqnarray}
where
\begin{eqnarray}
 {\cal\widetilde{O}}_2 &=&  (\overline{c} \gamma_{\mu}(1-\gamma_5)
\frac{\lambda_a}{2} c)
(\overline{d} \gamma^{\mu}(1-\gamma_5) \frac{\lambda_a}{2} b) \, ,  \nonumber \\
 {\cal\widetilde{O}}_3 &=&  (\overline{c} \gamma_{\mu}(1-\gamma_5)
\frac{\lambda_a}{2} c)
(\overline{d} \gamma^{\mu}(1-\gamma_5) \frac{\lambda_a}{2} b) \, , \nonumber\\
 {\cal\widetilde{O}}_5 &=&  (\overline{c} \gamma_{\mu}(1+\gamma_5)
\frac{\lambda_a}{2} c) (\overline{d} \gamma^{\mu}(1-\gamma_5)
\frac{\lambda_a}{2} b) \, ,
\end{eqnarray}
 and $\lambda^a$'s are $SU(3)$ Gell-Mann matrices.

The factorizable  matrix elements $\langle \chi_{c1}(p) \pi^0(-q) |
H_w | B(p-q) \rangle $    can be parameterized  as $\epsilon^* \cdot
q A_f$,
\begin{eqnarray}
 \epsilon^* \cdot q
A_f&=& \frac{G_F}{\sqrt{2}} \left\{ -V_{c b} V_{c d}^* \left[
C_2(\mu) + \frac{C_1(\mu)}{3} \right] \right.\nonumber\\
&&\left.+ V_{t b} V_{t d}^* \left [ C_3(\mu) + \frac{C_4(\mu)}{3}
-C_5(\mu)- \frac{C_6(\mu)}{3}\right ]\right\} \nonumber\\
&&\langle \chi_{c1}(p) | \overline{c} \gamma_{\mu} \gamma_5 c | 0
\rangle \langle \pi^0(-q)| \overline{d} \gamma^{\mu}  b | B(p-q)
\rangle \, .
\end{eqnarray}
 The $\chi_{c1}$ meson decay constant is defined
by  $ \langle \chi_{c1}(p) | \overline{c} (0) \gamma_{\mu}\gamma_5
c(0) |0\rangle = f_{\chi_{c1}} m_{\chi_{c1}}\epsilon_{\mu}^*$. The
$B-\pi^0$ form-factor can be parameterized as
\begin{eqnarray}
-\sqrt{2}\langle \pi^0(q)| \bar{d} \gamma_{\mu} b | B(P) \rangle &=&
f(p^2) (P+q)_{\mu} - [f(p^2) - f_0(p^2)]\frac{m_B^2 - m_\pi^2}{p^2}
p_{\mu}  \, ,
\end{eqnarray}
the  form-factors $f(p^2)$ and $f_0(p^2)$ can be estimated with the
light-cone sum rules approach, here we take the value
$f(m_{\chi_{c1}}^2) = 0.62$ \cite{PBall2004}.

 The concise expression for the
factorizable matrix elements can be written as
\begin{eqnarray}
 A_f&=&-G_Ff_{\chi_{c1}}m_{\chi_{c1}}f(m_{\chi_{c1}}^2)\left\{
 V_{c b} V_{c d}^* \left [ C_2(\mu) + \frac{C_1(\mu)}{3} \right ]- \right.\nonumber\\
&& \left.V_{t b} V_{t d}^*\left[ C_3(\mu)-C_5(\mu) +
\frac{C_4(\mu)-C_6(\mu)}{3}\right]  \right \} \, .
\end{eqnarray}

\section{ Light-cone QCD sum rules for the nonfactorizable hadronic matrix element
$\langle \chi_{c1} \pi^0 | {\cal \widetilde{O}}(0) | B \rangle $
}

In the following, we apply the approach developed in
Ref.\cite{Kh2001} for the $B \rightarrow \pi \pi$ channel to
estimate the contribution from the soft-gluon exchanges  in the
decay $B^0 \to \chi_{c1} \pi^0$. We write down the correlation
function $\Pi_{\rho}(p,q,k)$ firstly,
\begin{eqnarray}
\Pi_{\rho}(p,q,k) &=& i^2 \, \int d^4 x e^{-i(p-q)x} \int d^4 y
e^{i(p-k)y} \langle 0 | T \{ J_{\rho}(y) {\cal \widetilde{O}}(0)
J_5(x) \} | \pi^0(q) \rangle \, ,
\end{eqnarray}
where ${\cal \widetilde{O}}={\cal \widetilde{O}}_2$, the currents
$J_{\rho} = \bar{c} \gamma_{\rho}\gamma_5 c$ and $J_5 = m_b \bar{b}
i \gamma_5 d$ interpolate the mesons $\chi_{c1}$
 and $B$,  respectively.

The correlation function $\Pi_{\rho}(p,q,k)$ can be calculated  by
the operator product expansion approach  near the light-cone
$x^2\sim y^2\sim (x-y)^2\sim 0$ in perturbative QCD theory. It is
function of three independent momenta chosen to be $q$, $p-k$ and
$k$. We introduce the unphysical momentum $k$ in order to avoid that
the $B$ meson has the same four-momentum before ($p-q$) and after
the decay ($P$), and thus avoid a continuum of light contributions
in the dispersion relation in the $B$-channel. The independent
kinematical invariants can be taken as $(p-q)^2$, $(p-k)^2$, $q^2$ ,
$k^2$,  $P^2 = (p-k-q)^2$ and $p^2$. We set $k^2 =0$ and take $q^2 =
m_\pi^2 =0$, neglecting the small corrections of  order
$O(m_\pi^2/m_B^2)$. The momentum $p^2$  is kept undefined for the
moment  in order to   make the derivation of the sum rules without
restriction. Its value is going to be set later in this section, and
chosen to be $p^2 =m_{\chi_{c1}}^2$. The values of $(p-k)^2$,
$(p-q)^2$ and $P^2$ should be spacelike and large in order to stay
far away from the hadronic thresholds in the $B$ and $\chi_{c1}$
channels. All together, we have
\begin{eqnarray}
q^2 = k^2 = 0, \; p^2 = m_{\chi_{c1}}^2, \; |(p-k)|^2 \gg
\Lambda_{QCD}, |(p-q)|^2 \gg\Lambda_{QCD},  |P|^2 \gg \Lambda_{QCD}
\,\, .\nonumber
\end{eqnarray}

 The correlation function $\Pi_\rho(p,q,k)$ can be decomposed  as
\begin{eqnarray}
\Pi_\rho(p,q,k)&=& (p-k)_\rho \Pi + q_\rho \widetilde{\Pi}_1 +
k_\rho \widetilde{\Pi}_2 + \epsilon_{\rho \alpha\beta\sigma}q^\alpha
p^\beta k^\sigma \widetilde{\Pi}_3\, ,
\end{eqnarray}
due to  Lorentz covariance.

 According to the basic assumption of
quark-hadron duality in the QCD sum rules \cite{SVZ,PRT85}, we
insert  a complete set of intermediate states with the same quantum
numbers as the current operators $J_\mu(y)$ and $J_5(x) $ into the
correlation function $\Pi_\rho(p,q,k)$ to obtain the hadronic
representation. After isolating the pole terms of the ground state
mesons $\chi_{c1}$ and $B$, we get the following result,
\begin{eqnarray}
\Pi_{\rho} &=& \frac{\langle 0 | J_{\rho}(0) |\chi_{c1} (p-k)
\rangle}{m^2_{\chi_{c1}}-(p-k)^2} \langle \chi_{c1}
(p-k)|{\cal\widetilde{ O}}(0)|B(p-q)\rangle | \pi^0(q)
\rangle\frac{\langle B(p-q)|  J_5(0) | 0 \rangle}{m^2_{B}-(p-q)^2}
\nonumber \\
&&   +\cdots,  \nonumber\\
&=&\frac{f_{\chi_{c1}}m_{\chi_{c1}}\epsilon_{\rho}}{m^2_{\chi_{c1}}-(p-k)^2}\frac{f_B
m^2_B}{m^2_{B}-(p-q)^2}
\langle \chi_{c1} (p-k)\pi^0(-q)|{\cal \widetilde{O}}(0)|B(p-q)\rangle+\cdots , \nonumber \\
 &=&\frac{f_{\chi_{c1}}m_{\chi_{c1}}}{m^2_{\chi_{c1}}-(p-k)^2}\frac{f_B
m^2_B}{m^2_{B}-(p-q)^2}
  q_\alpha \left[-g_{\alpha\rho}+\frac{(p-k)_\alpha (p-k)_\rho}{(p-k)^2} \right]A_n +\cdots , \nonumber \\
\end{eqnarray}
where we have used the definition $\langle \chi_{c1}
(p-k)\pi^0(-q)|{\cal \widetilde{O}}(0)|B(p-q)\rangle=\epsilon^*
\cdot q A_n$, and do not show the contributions from the higher
resonances and continuum states above the corresponding thresholds
explicitly, they can be written in terms of dispersion integrals and
the spectral density  can be approximated by the quark-hadron
duality ansatz.

Now we carry out the operator product expansion near the light-cone
to obtain the    representation at the level of quark-gluon degrees
of freedom. Let us write down the propagator of a massive quark in
the external gluon field in the Fock-Schwinger gauge firstly
\cite{BBKR1995},
\begin{eqnarray}
S_{ij}(x,y)& =&i \int\frac{d^4k}{(2\pi)^4}e^{-ik(x-y)}\Bigg\{
\frac{\not\!k +m}{k^2-m^2} \delta_{ij} -\int\limits_0^1 dv\,  g_s \,
G^{\mu\nu}_{ij}(vx+(1-v)y)  \nonumber
\\
& &  \Big[ \frac12 \frac {\not\!k +m}{(k^2-m^2)^2}\sigma_{\mu\nu} -
\frac1{k^2-m^2}v(x-y)_\mu \gamma_\nu \Big]\Bigg\}\, ,
\end{eqnarray}
where $G^{\mu \nu }_a$ is the gluonic field strength, $g_s$ denotes
the strong coupling constant.

Substituting the above $b$ and $c$ quark propagators  into the
correlation function $\Pi_\rho$, and completing the corresponding
integrals,  we can obtain the hadronic spectral density at the level
of quark-gluon degrees of freedom. In calculation, the following
three-particle $\pi^0$ light-cone  distribution amplitudes are
useful,
\begin{eqnarray}
-\sqrt{2} \langle 0 |\bar{d}(0) \sigma_{\mu \nu} \gamma_5 G_{\alpha
\beta}(v y) d(x) | \pi^{0}(q) \rangle &=&
i f_{3 \pi} \left [ q_{\alpha}q_{\mu}g_{\beta \nu} +q_{\beta}q_{\nu}g_{\alpha \mu}- q_{\beta}q_{\mu}g_{\alpha \nu}- \right .  \nonumber \\
& &     \left . q_{\alpha}q_{\nu}g_{\beta \mu} \right ] \int {\cal
D}\alpha_i \phi_{3 \pi}(\alpha_i) e^{-i q(x \alpha_1 + y v
\alpha_3)} \, , \nonumber \\
-\sqrt{2} \langle 0 |\bar{d}(0) i\gamma_{\mu} \tilde{G}_{\alpha
\beta}(v y) d(x) |\pi^{0}(q) \rangle &=& f_{\pi}q_{\mu}
\frac{q_{\alpha} x_{\beta} - q_{\beta}x_{\alpha}}{q x}  \int {\cal
D}\alpha_i \widetilde{\phi}_{\parallel}(\alpha_i) e^{-i q(x \alpha_1
+ y v \alpha_3)} +\nonumber\\
 & &   f_{\pi}(g_{\mu
\alpha}^{\perp}q_{\beta} - g_{\mu \beta}^{\perp}q_{\alpha}) \int
{\cal D}\alpha_i \widetilde{\phi}_{\perp}(\alpha_i) e^{-i q(x
\alpha_1 + y v
\alpha_3)} \, , \nonumber \\
-\sqrt{2} \langle 0 |\bar{d}(0) \gamma_{\mu} \gamma_5 {G}_{\alpha
\beta}(v y) d(x) |\pi^{0}(q) \rangle &=& f_{\pi} q_{\mu}
\frac{q_{\alpha} x_{\beta} - q_{\beta}x_{\alpha}}{q x}  \int {\cal
D}\alpha_i {\phi}_{\parallel}(\alpha_i)
e^{-i q(x \alpha_1 + y v \alpha_3)} +\nonumber \\
& &  f_{\pi}(g_{\mu \alpha}^{\perp}q_{\beta} - g_{\mu
\beta}^{\perp}q_{\alpha}) \int {\cal D}\alpha_i
{\phi}_{\perp}(\alpha_i) e^{-i q(x \alpha_1 + y v \alpha_3)} \, ,
\nonumber \\
\end{eqnarray}
where $\tilde{G}_{\alpha \beta} = \frac{1}{2} \epsilon_{\alpha \beta
\rho \sigma} G^{\rho \sigma}$, $ {\cal D} \alpha_i = d\alpha_1
d\alpha_2 d\alpha_3 \delta( 1 - \alpha_1 - \alpha_2 - \alpha_3)$,
$g_{\alpha \beta}^{\perp} = g_{\alpha \beta} - \frac{x_{\alpha}
q_{\beta} + x_{\beta} q_{\alpha}}{qx}$. The twist-3 and twist-4
light-cone distribution amplitudes can be parameterized as
\begin{eqnarray}
\phi_{3\pi}(\alpha_i) &=& 360 \alpha_1 \alpha_2 \alpha_3^2 \left
[ 1+\frac{\omega_3}{2}(7\alpha_3-3)\right] \, , \nonumber \\
\phi_{\perp}(\alpha_i) &=& 30
\delta^2(\alpha_1-\alpha_2)\alpha_3^2\left [ \frac{1}{3} + 2
\epsilon  (1 - 2 \alpha_3) \right ]  \, , \nonumber \\
\phi_{\parallel}(\alpha_i) &=& 120 \delta^2 \epsilon
(\alpha_1-\alpha_2) \alpha_1 \alpha_2 \alpha_3  \, ,\nonumber\\
\widetilde{\phi}_{\perp}(\alpha_i) &=& 30 \delta^2 \alpha_3^2 ( 1 -
\alpha_3) \left [ \frac{1}{3} + 2 \epsilon
(1 - 2\alpha_3) \right ] \, ,\nonumber \\
\widetilde{\phi}_{\parallel}(\alpha_i) &=& -120 \delta^2 \alpha_1
\alpha_2 \alpha_3 \left [ \frac{1}{3} + \epsilon  (1 - 3 \alpha_3)
\right ] \, ,
\end{eqnarray}
the nonperturbative parameters in the light-cone distribution
amplitudes can be estimated with the QCD sum rules
\cite{BF1990,CZ1984,BallJHEP}.

After carrying out the operator product expansion near the
light-cone, we obtain the following expression for the correlation
function $\Pi$ \footnote{For technical details, one can consult
Ref.\cite{Soft4}.},
\begin{eqnarray}
\Pi &=&  \frac{m_b f_{3 \pi}}{2 \pi^2} \int_0^1 dv \int_0^1 dt\int
{\cal D}\alpha_i \frac{\phi_{3\pi}(\alpha_i)q \cdot (p-k)\left [ tv
q \cdot k + 2 (1-t)(1-v) q \cdot p \right ] } {\left\{m_b^2 - [p - q
(1-\alpha_1)]^2\right\}\left\{\widetilde{m}_c^2 - (p-k-v \alpha_3
q)^2\right\}}  \nonumber \\
&&+\frac{m^2_b f_{ \pi}}{4 \pi^2} \int_0^1 dv \int_0^1 dt\int {\cal
D}\alpha_i \frac{q\cdot(p-k)} {\left\{m_b^2 - [p - q
(1-\alpha_1)]^2\right\}\left\{\widetilde{m}_c^2 - (p-k-v \alpha_3
q)^2\right\}}  \nonumber \\
&&\left\{ (1-2t)\left[2(1-v)\phi_\perp(\alpha_i)
-(1-2v)\phi_\parallel(\alpha_i)\right]
+2(1-v)\widetilde{\phi}_\perp(\alpha_i)-\widetilde{\phi}_\parallel(\alpha_i)\right\}\, ,\nonumber\\
\end{eqnarray}
where $\widetilde{m}_c^2=\frac{m_c^2}{t(1-t)}$.

  In order to suppress the contributions from the high resonances and
continuum states, we  perform n-th derivative with respect to the
momentum $(p-k)^2$ to obtain stable n-th moment sum rules and  Borel
transform with respect to the momentum $(p-q)^2$ in the $B$-channel
to obtain the Borel sum rules,
 then match  with Eq.(10),
finally we obtain the sum rule for the nonfactorizable soft matrix
element,
\begin{eqnarray}
A_n &=& \frac{2m_{\chi_{c1}}}{f_B f_{\chi_{c1}} m^2_B
\left[m_{\chi_{c1}}^2-P^2\right]} \int_{4 m^2_c}^{s_0}d s \left\{
\frac{f_{3\pi}m_b}{4\pi^2}\int_{t_i}^{t_f}dt \int_{\alpha_0}^1
d\alpha
\int_{\alpha_c}^{\alpha} d \beta  \right. \nonumber\\
&& \left.\phi_{3\pi}(1-\alpha,\alpha-\beta,\beta)\left[
\frac{t\alpha_c}{2\beta}(P^2-s-\frac{m_b^2-m_{\chi_{c1}}^2}{\alpha})
-(1-t)(1-\frac{\alpha_c}{\beta})\frac{m_b^2-m^2_{\chi_{c1}}
}{\alpha} \right] \right.  \nonumber \\
&& +\frac{f_\pi m_b^2 }{8\pi^2}\int_{t_i}^{t_f}dt \int_{\alpha_0}^1
d\alpha \int_{\alpha_c}^{\alpha} d \beta
\left[2(1-\frac{\alpha_c}{\beta})\widetilde{\phi}_{\perp}
 -\widetilde{\phi}_{\parallel}+\right.  \nonumber\\
&&\left. \left.(1-2t)\left[2(1-\frac{\alpha_c}{\beta})\phi_\perp -(1-\frac{2\alpha_c}{\beta})\phi_\parallel  \right] \right](1-\alpha,\alpha-\beta,\beta)\right\}\nonumber\\
 &&\exp\left(\frac{\alpha m_B^2 +m^2_{\chi_{c1}}(1-\alpha)-m^2_b}{\alpha M^2 }\right) \left(\frac{m^2_{\chi_{c1}}+Q_0^2}{s+Q_0^2}\right)^{n+1}\frac{1}{\alpha\beta }\,
 ,
\end{eqnarray}
where
\begin{eqnarray}
t_i=\frac{1}{2} \left(1-\sqrt{1-\frac{4m_c^2}{s}} \right) \, , & \ \ \ & t_f=\frac{1}{2} \left(1+\sqrt{1-\frac{4m_c^2}{s}} \right)\, , \nonumber \\
\alpha_c=\frac{t(1-t)s-m_c^2}{t(1-t)(s-P^2)} \, , & \ \ \ &
\alpha_0=\frac{m_b^2-m^2_{\chi_{c1}}}{s_B-m^2_{\chi_{c1}}} \, ,
\end{eqnarray}
 and $Q_0^2=4m_c^2\xi$. In  calculation, $P^2$ is chosen to be
large space-like squared momentum ($|P^2|\sim m_b^2$) in order to
stay far away from the hadronic thresholds,  the value of $\alpha_c$
is a small positive quantity but not always small enough to be
safely
  neglected, we  perform the following approximation for the $\beta$
integral,
\begin{eqnarray}
\int_{\alpha_c}^{\alpha} d\beta
G(s,x,\alpha,\beta)=\left\{\int^{\alpha}_{0}
-\int^{\alpha_c}_{0}\right\} d\beta G(s,x,\alpha,\beta) \, ,
\end{eqnarray}
here $G$ is an abbreviation for the integral functions and can be
written as
\begin{eqnarray}
G(s,x,\alpha,\beta)&=&A\phi_{3\pi}(1-\alpha,\alpha-\beta,\beta)
+B\widetilde{\phi}_{\parallel}(1-\alpha,\alpha-\beta,\beta)
+C\widetilde{\phi}_{\perp}(1-\alpha,\alpha-\beta,\beta) \nonumber \\
&&+ D\phi_{\parallel}(1-\alpha,\alpha-\beta,\beta)
+E\phi_{\perp}(1-\alpha,\alpha-\beta,\beta)\, , \nonumber
\end{eqnarray}
$A,B,C,D,E$ are formal notations. We can expand the light-cone
distribution amplitudes $\phi_{3\pi}$, $
\widetilde{\phi}_{\parallel}$, $\widetilde{\phi}_{\perp}$,
$\phi_{\parallel}$ and $\phi_{\perp}$ in terms of Taylor series of
$\beta$, for example,
\begin{eqnarray}
\phi_{3\pi}(1-\alpha,\alpha-\beta,\beta)&=&\phi_{3\pi}(1-\alpha,\alpha-\beta,\beta)|_{\beta=0}+\frac{\partial}{\partial
\beta }\phi_{3\pi}(1-\alpha,\alpha-\beta,\beta)|_{\beta=0}\beta\nonumber\\
 &&+\frac{1}{2}\frac{\partial^2}{\partial \beta^2
}\phi_{3\pi}(1-\alpha,\alpha-\beta,\beta)|_{\beta=0}\beta^2+\cdots
\, ,
\end{eqnarray}
 and  continue $P^2$ into the
timelike region analytically, $P^2=m^2_B$, then complete the
integral $\int^{\alpha_c}_{0} d\beta G(s,x,\alpha,\beta)$. This
procedure ensures the  disappearance of the unphysical momentum $k$
from the ground state contribution and enables the extraction of the
physical matrix element due to the simultaneous conditions, $P^2 =
m_B^2$ and $(p-q)^2 = m_B^2$.

The explicit expression for the physical hadronic matrix element  is
 lengthy due to the re-summation of all the Taylor series of $\beta$,  here we show only the leading terms explicitly,
\begin{eqnarray}
A_n &=& \frac{2m_{\chi_{c1}}}{f_B f_{\chi_{c1}} m^2_B
\left[m_{\chi_{c1}}^2-m_B^2\right]} \int_{4 m^2_c}^{s_{0}}d s
\left\{ \frac{f_{3\pi}m_b}{4\pi^2}\int_{t_i}^{t_f}dt
\int_{\alpha_0}^1 d\alpha
\int_{0}^{\alpha} d \beta  \right. \nonumber\\
&& \phi_{3\pi}(1-\alpha,\alpha-\beta,\beta)\left[
\frac{t\alpha_c}{2\beta}(m_B^2-s-\frac{m_b^2-m_{\chi_{c1}}^2}{\alpha})
-(1-t)(1-\frac{\alpha_c}{\beta})\frac{m_b^2-m^2_{\chi_{c1}}
}{\alpha}\right]\nonumber\\
&& +\frac{f_\pi m_b^2 }{8\pi^2}\int_{t_i}^{t_f}dt \int_{\alpha_0}^1
d\alpha \int_{0}^{\alpha} d \beta
\left[2(1-\frac{\alpha_c}{\beta})\widetilde{\phi}_{\perp}
 -\widetilde{\phi}_{\parallel}+\right.  \nonumber\\
&&\left. \left.(1-2t)\left[2(1-\frac{\alpha_c}{\beta})\phi_\perp -(1-\frac{2\alpha_c}{\beta})\phi_\parallel  \right] \right](1-\alpha,\alpha-\beta,\beta)\right\}\nonumber\\
 &&\exp\left(\frac{\alpha m_B^2 +m^2_{\chi_{c1}}(1-\alpha)-m^2_b}{\alpha M^2 }\right)
 \left(\frac{m^2_{\chi_{c1}}+Q_0^2}{s+Q_0^2}\right)^{n+1}\frac{1}{\alpha\beta
 } \nonumber\\
 &&- \frac{2m_{\chi_{c1}}}{f_B f_{\chi_{c1}} m^2_B
\left[m_{\chi_{c1}}^2-m_B^2\right]} \int_{4 m^2_c}^{s_{0}}d s
\left\{ \frac{f_{3\pi}m_b}{4\pi^2}\int_{t_i}^{t_f}dt
\int_{\alpha_0}^1 d\alpha
\int_{0}^{\alpha_c} d \beta  \right. \nonumber\\
&& \phi_{3\pi}(1-\alpha,\alpha-\beta,\beta)\mid_{\beta=0}\left[
\frac{t\alpha_c}{2\beta}(m_B^2-s-\frac{m_b^2-m_{\chi_{c1}}^2}{\alpha})
-\right.\nonumber\\
&&\left.(1-t)(1-\frac{\alpha_c}{\beta})\frac{m_b^2-m^2_{\chi_{c1}}
}{\alpha} \right]   \nonumber \\
&& +\frac{f_\pi m_b^2 }{8\pi^2}\int_{t_i}^{t_f}dt \int_{\alpha_0}^1
d\alpha \int_{0}^{\alpha_c} d \beta
\left[2(1-\frac{\alpha_c}{\beta})\widetilde{\phi}_{\perp}
 -\widetilde{\phi}_{\parallel}+\right.  \nonumber\\
&&\left. \left.(1-2t)\left[2(1-\frac{\alpha_c}{\beta})\phi_\perp -(1-\frac{2\alpha_c}{\beta})\phi_\parallel
 \right] \right](1-\alpha,\alpha-\beta,\beta)\mid_{\beta=0}\right\}\nonumber\\
 &&\exp\left(\frac{\alpha m_B^2 +m^2_{\chi_{c1}}(1-\alpha)-m^2_b}{\alpha M^2 }\right)
 \left(\frac{m^2_{\chi_{c1}}+Q_0^2}{s+Q_0^2}\right)^{n+1}\frac{1}{\alpha\beta
 }+\cdots\, \, .
 \end{eqnarray}
 In performing the $\beta$ integral $\int_0^{\alpha_c}$, we need only the
values of the light-cone distribution amplitudes $\phi_{3\pi}$, $
\widetilde{\phi}_{\parallel}$, $\widetilde{\phi}_{\perp}$,
$\phi_{\parallel}$, $\phi_{\perp}$ and their derivations at zero
momentum fraction i.e. $\beta=0$, there are no problems with
negative partons (quarks and gluons) momentum fractions. The
analytical continuation of
 $P^2$ to its positive value ends up with an unavoidable
theoretical uncertainty, if only a few terms of the Taylor series
are taken. However, with the re-summation to all orders of $\beta$
in Eq.(19), the  assumption of quark-hadron duality is still
applicable  in  the case of heavy meson final states.

 \section{Numerical results }

The input parameters   are taken as $V_{cd}=-0.230$,
$V_{cb}=41.2\times 10^{-3}$, $V_{tb}=1.0$, $V_{td}=8.1 \times
10^{-3}$, $m_{\chi_{c1}} = 3.511\, \rm{GeV}$, $m_B = 5.2795\, \rm{
GeV}$ \cite{PDG},  $f_B = 0.18 \, \rm{GeV} $, $s_{B} = 35 \pm 2\,
{\rm GeV}^2$ \cite{Kho9801},
 $f_{\chi_{c1}} = 0.335 \,
\rm{GeV}$ \cite{PRT78}, $s_0 = (16.0 \pm 0.5) \, {\rm GeV}^2$
\cite{PRT85},  $f_\pi = 0.13 \, \rm{GeV}$, $m_b = (4.7 \pm 0.1)\,
\rm{ GeV}$, $m_c =(1.35 \pm 0.05)\, \rm{GeV}$,
$f_{3\pi}=(0.45\pm0.15)\times 10^{-2}\,\rm{GeV}^2$,
$\omega_3=-1.5\pm0.7$, $\delta^2 = (0.18\pm0.06)\, \rm{GeV}^2$,
$\varepsilon=\frac{21}{8}(0.2\pm0.1)$ \cite{BF1990,CZ1984,BallJHEP}
at the energy scale about $\mu=1\, \rm{GeV}$.

The parameters $n$ and $M^2$ must be carefully chosen to warrant the
high resonances  and continuum states to be  suppressed and obtain a
reliable perturbative QCD calculation. The stable region for the
Borel parameter $M^2$ is found in the interval $M^2 = 11 \pm 2\,
{\rm GeV}^2$, which is known from the  $B$ channel QCD sum rules
\cite{Kho9801}. In the charmonium channels, we usually perform n-th
derivative and take n-th moment sum rules to satisfy the stability
criteria  \cite{PRT85}, the ideal interval is $n=4-6$.    The
parameter $\xi$ is usually allowed to take values larger than 1 for
the $P$-wave charmonia, we observe the best interval is
$\xi=1.8-2.6$.

The nonfactorizable soft contributions come from the three-particle
twist-3 and twist-4 $\pi^0$    light-cone distribution amplitudes,
however, present knowledge about those distribution amplitudes is
rather poor. The uncertainties of the nonperturbative parameters
$f_{3\pi}$, $\omega_3$, $\delta^2 $ and $\varepsilon$ are  large,
about $(33-50)\%$, the  nonfactorizable soft contributions $A_n$ can
be simplified into the following form,
\begin{eqnarray}
A_n&=& y_1 f_{3\pi}+y_2 f_{3\pi}\omega_3+y_3\delta^2
+y_4\delta^2\varepsilon \, ,
\end{eqnarray}
where the $y_i$ are numerical coefficients not shown explicitly. The
uncertainties origin from the nonperturbative parameters $f_{3\pi}$,
$\omega_3$, $\delta^2 $ and $\varepsilon$ are rather large, even out
of control. We take the same treatment as
Refs.\cite{Soft4,Soft5,Soft6,Soft7,Soft10},
 and neglect the corresponding uncertainties,  it weakens the
predictive ability. The uncertainties origin from other parameters
($s_B$, $s_0$, $m_b$, $m_c$, $n$, $\xi$, etc) are less than $22\%$,
we take into account  them  with the formula
$\Delta=\sqrt{\sum_i\left(\frac{\partial f}{\partial x_i}\right)^2
(x_i-\bar{x}_i)^2}$, where the $f$ denote  the nonfactorizable soft
contributions, the $x_i$ denote the input parameters.

Taking into account the next-to-leading order Wilson coefficients
calculated in the naive dimensional regularization scheme
\cite{Buras} for $\mu = \overline{m_b}(m_b) = 4.40 {\rm \, GeV}$ and
$\Lambda_{\overline{\rm MS}}^{(5)} = 225\, {\rm MeV}$,
\begin{eqnarray}
\qquad C_1(\overline{m_b}(m_b)) = 1.082 \, , \qquad C_2(\overline{m_b}(m_b)) = -0.185 \, ,\qquad C_3(\overline{m_b}(m_b)) = 0.014 \, , \nonumber\\
\qquad C_4(\overline{m_b}(m_b)) = -0.035 \, ,\qquad
C_5(\overline{m_b}(m_b)) = 0.009 \, ,\qquad C_6(\overline{m_b}(m_b))
= -0.041 \, ,
\end{eqnarray}
  finally  we obtain the numerical ratio  $R$ between  the
contributions from the nonfactorizable and factorizable hadronic
matrix elements, which is shown in Fig.1,
\begin{eqnarray}
R&=&\frac{\sqrt{2} C_1 (\mu)  \langle \chi_{c1} (p)\pi^0(-q)|{\cal
{\widetilde{O}}}(0)|B(p-q)\rangle} {-\left[C_2(\mu)+\frac{C_1
(\mu)}{3}\right]f_{\chi_{c1}} m_{\chi_{c1}}
f(m^2_{\chi_{c1}})}=0.68^{+0.20}_{-0.16}  \,\, .
\end{eqnarray}

The nonfactorizable soft contributions are rather large  and they
must be included in analyzing the branching fraction. The total
amplitudes lead to the branching fraction,
\begin{eqnarray}
 Br(B^0\to \chi_{c1}\pi^0)= 1.41^{+0.35}_{-0.25} \times 10^{-5} \, ,
\end{eqnarray}
which is in agreement with the experimental  data $ (1.12\pm 0.25\pm
0.12)\times 10^{-5}$  marginally \cite{Belle0809}.

The factorizable  contributions of the operator  ${\cal O}_2(0)$ can
be analyzed in the same way with the correlation function
$\widehat{\Pi}_{\rho}(p,q,k)$,
\begin{eqnarray}
\widehat{\Pi}_{\rho} &=& i^2 \, \int d^4 x e^{-i(p-q)x} \int d^4 y
e^{i(p-k)y} \langle 0 | T \{ J_{\rho}(y) {\cal O}_2(0) J_5(x) \} |
\pi^0(q) \rangle  \nonumber\\
&=&  \int d^4 x e^{-i(p-q)x}  \langle 0 | T [
\overline{d}(0)\gamma^\mu b(0) J_5(x) ] |
\pi^0(q) \rangle\int d^4 y e^{i(p-k)y} \langle 0 | T [ J_{\rho}(y) J_\mu (0) ] |0 \rangle \nonumber\\
&\propto& (p-k)_\rho  f\left((p-k)^2\right)\times
f_{\chi_{c1}}^2+\cdots \, .
\end{eqnarray}
The contributions of the soft gluons can be absorbed into the $B \to
\pi^0$ form-factor $f\left((p-k)^2\right)$ or the decay constant
$f_{\chi_{c1}}$, and differ from the  sum rules for the operator
${\cal {\widetilde{O}}}(0)$ greatly, where the contributions of the
soft gluons are nonfactorizable, so we are free of double-counting.

\begin{figure}
 \centering
 \includegraphics[totalheight=6cm,width=8cm]{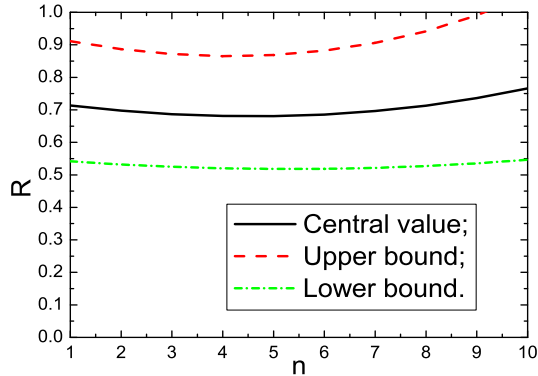}
    \caption{ The ratio $R$ between  the contributions from the
nonfactorizable and factorizable hadronic  matrix elements. }
\end{figure}

 \section{Conclusion}
In this article, we calculate  the contributions from the
nonfactorizable soft hadronic matrix element  $\langle \chi_{c1}
(p)\pi^0(-q)|{\cal {\widetilde{O}}}(0)|B(p-q)\rangle$
   to the  decay
$B^0\rightarrow \chi_{c1} \pi^0$ with  the light-cone QCD sum rules.
The numerical results show that its contribution is rather large and
should not be neglected. The total amplitudes lead to a branching
fraction which is in agreement with the experimental data
marginally.

\section*{Acknowledgements}
This  work is supported by National Natural Science Foundation,
Grant Number 10775051, and Program for New Century Excellent Talents
in University, Grant Number NCET-07-0282.

\end{document}